\documentclass[runningheads]{llncs}
\usepackage[T1]{fontenc}
\usepackage{graphicx}
\usepackage{amsfonts,amsmath,amssymb}
\usepackage{bm} 
\def \vect#1{\bm{#1}}
\def \matr#1{\bm{#1}}

\usepackage{tikz}
\usetikzlibrary{arrows, positioning,arrows.meta}
\usetikzlibrary{backgrounds}
\usetikzlibrary{calc}
\usepackage[abs]{overpic}

\begin{document}
\title{Adjoint-based optimization with quantized local reduced-order models for spatiotemporally chaotic systems}
\titlerunning{Adjoint-based optimization with quantized local reduced-order models}
%
\author{Defne Ege Ozan\inst{1}\orcidID{0009-0006-4208-9981} \and
Antonio Colanera\inst{2}\orcidID{0000-0002-4227-1162} \and
Luca Magri\inst{1,3}\orcidID{0000-0002-0657-2611}}
\authorrunning{D.E. Ozan, A. Colanera, L. Magri}
%
\institute{Department of Aeronautics, Imperial College London, SW7 2AZ, London, UK \email{l.magri@imperial.ac.uk}\and International School for Advanced Studies (SISSA), \\Via Bonomea 265, Trieste 34136, Italy \and Politecnico di Torino, DIMEAS, Corso Duca degli Abruzzi, 24 10129 Torino, Italy}
\maketitle              
\begin{abstract}
We introduce a computationally efficient and accurate reduced-order modelling approach for the optimization of spatiotemporally chaotic systems. The proposed method combines quantized local reduced-order modelling with adjoint-based optimization. We employ the methodology in a variational data assimilation problem for the chaotic Kuramoto-Sivashinsky equation and show that it successfully reconstructs the full trajectory for up to 0.25 Lyapunov times given full state measurements at the final time. The proposed algorithm provides $\times 3.5$ speed-up when compared to the full-order model. The proposed method opens up new possibilities for the reduced-order modelling of spatiotemporally chaotic systems.
\end{abstract}
\section{Introduction}
Fluid systems that are of scientific and engineering interest are typically described by partial differential equations (PDEs), which often have spatiotemporally chaotic solutions \cite{holmes_turbulence_1996}. Accurate solutions of these PDEs typically require fine discretizations that yield high-dimensional state representations in order to resolve the underlying multi-scale dynamics. The resulting high computational cost for the simulation of these systems motivates the development of reduced-order modelling (ROM) approaches, which aim to discover low-dimensional models that can be utilized to predict the system behaviour with some approximation error \cite{quarteroni_certified_2011}. In dissipative systems, which are common in fluids, the ROM approach becomes viable, as the trajectories are attracted to a low-dimensional manifold. 

ROMs are especially appealing in gradient-based optimization, where repeated forward simulations are required to reach convergence. Adjoint methods are central in this context. The adjoint formulation enables the computation of gradients of an objective functional with respect to many design parameters at a cost that is essentially independent of the number of parameters ~e.g.,~\cite{giles_introduction_2000,magri2019AdjointMethodsDesign,Ozan2024}, unlike tangent-linear or finite-difference approaches whose cost scales with the parameter dimension.  In data assimilation, an alternative class of approaches is based on sequential filtering, most notably the ensemble Kalman filter (EnKF); a general EnKF formalism is presented in \cite{Novoa2024a}.

In this paper, we leverage the recently developed quantized local reduced-order models (ql-ROMs) \cite{colanera_quantized_2025} for gradient-based optimization, which overcomes the challenge of designing a single global ROM for the entire state space. In this approach, the manifold is quantized via unsupervised clustering, and intrusive local ROMs are built for each cluster. The switching between the local models is enabled via a change of basis using assignment functions. We derive the adjoint of the ql-ROM, where a similar coordinate transformation exists between the local adjoint variables. We demonstrate the application of this method in a variational data assimilation problem of a prototypical spatiotemporally chaotic system, where the objective is to infer the initial condition of a trajectory given observations. 

The remainder of the paper is organized as follows. In Section \ref{sec:methodology}, we present the proposed methodology. Specifically, Section \ref{sec:qlrom} summarizes the ql-ROM algorithm, Section \ref{sec:adjoint} derives the adjoint of the ql-ROM and Section \ref{sec:var_da} embeds it within a gradient-based optimization framework for variational data assimilation problems. The methodology is demonstrated on the spatiotemporally chaotic solutions of the Kuramoto-Sivashinsky (KS) equation in Section \ref{sec:application_ks}. Section \ref{sec:conclusion} concludes the paper.

\section{Methodology}\label{sec:methodology}
\begin{figure}
\centering

    \includegraphics[trim= 8cm 5cm 1.5cm 6cm,clip,width=\linewidth]{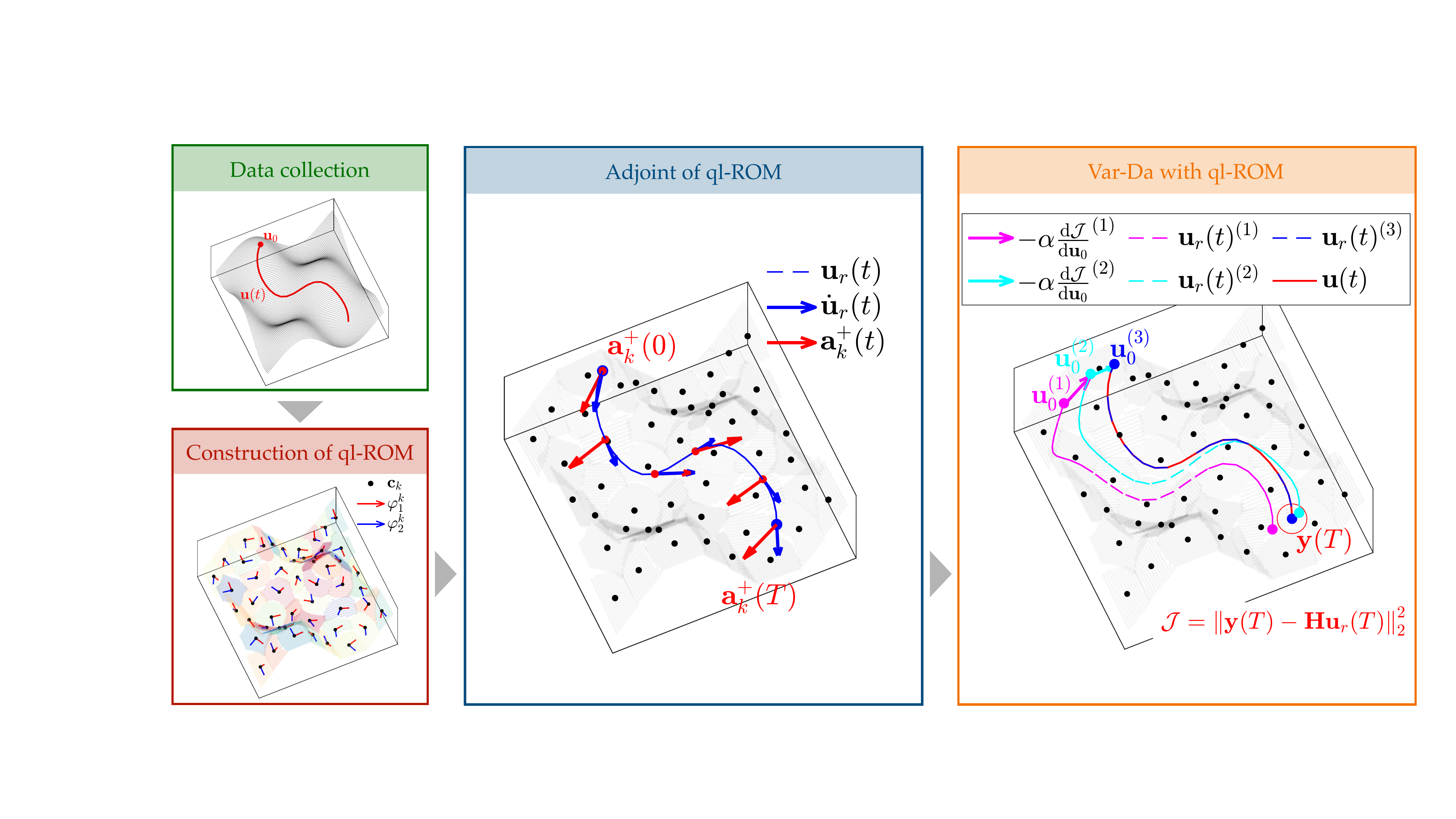}
\caption{
Overview of adjoint-based optimization with quantized local reduced-order models (ql-ROMs): (1) data collection; (2) phase-space quantization and local basis/model construction; (3) ql-ROM adjoint integrated backward in time with jump/coordinate-transformation at cluster switches; (4) variational data assimilation using the ql-ROM direct--adjoint loop to update the initial condition.
}\label{fig:localROM}
\end{figure}

In this section, we outline the proposed workflow to enable adjoint-based optimization with quantized local reduced-order models (ql-ROMs). The approach consists of three building blocks: the construction of local intrusive ROMs on a quantized state space, the derivation of the corresponding adjoint with appropriate jump/coordinate transformations at cluster switches, and the integration of this adjoint within a variational data assimilation loop to update the initial condition. The overall workflow is shown in Figure \ref{fig:localROM}.

\subsection{Quantized reduced-order models (ql-ROMs)}\label{sec:qlrom}
We consider the spatial discretization of a PDE which results in the following system of ordinary differential equations 
\begin{equation}
     \dot{\vect{u}} = \vect{f}(\vect{u}), \quad \vect{u} \in \mathbb{R}^{N}, \quad \vect{u}(t = 0) = \vect{u}_0
\end{equation}
where $\vect{u}$ is the state vector,  $\dot{\vect{u}}= d\vect{u}/dt$ is its time derivative, $\vect{f}$ is the nonlinear operator that describes the evolution of the state vector in time, and $N$ the dimension of the full-order state. Given a dataset of $M$ snapshots $\{\vect{u}(t_m)\}_{m=0}^{M}$, where $t_m = m\Delta t$ with $\Delta t$ being the time step, the method first partitions this data into $K$ clusters with centroids $\vect{c}_k \in \mathbb{R}^{N}$, $k = 1,2,\dots,K$. Here, we employ K-means for this purpose, which is an unsupervised machine learning method for clustering data based on a distance metric, e.g., the Euclidean distance from the centroid. The cluster affiliation function then assigns a snapshot to the closest cluster using this metric, where the centroid acts as the mean of the cluster. In the next step, we project $\vect{u}$ on a basis defined by a centroid $c_k$ and orthonormal basis vectors $\matr{V}_k \in \mathbb{R}^{N \times r_k}$ such that
\begin{equation}
    \vect{u} = \vect{c}_k + \matr{V}_k\vect{a}_k,
\end{equation}
where $\vect{a}_k \in \mathbb{R}^{r_k}$ are the coordinates in this new basis. The objective of reduced-order modelling is to build an accurate model with $r_k \ll N$. (When we set the number of clusters $K = 1$, this formulation becomes equivalent to the global modelling approach.) We determine
\begin{equation}
    \vect{a}_k = \matr{V}_k^\top(\vect{u}-\vect{c}_k).
\end{equation}
Here, we employ proper orthogonal decomposition (POD) \cite{holmes_turbulence_1996} to obtain the local bases $\matr{V}_k$ in each cluster. Using Galerkin projection of the dynamics onto the chosen basis vector, we obtain a model for the time evolution of $\vect{a}_k$ for each cluster
\begin{equation}\label{eq:dyn_a}
    \dot{\vect{a}}_k = \vect{g}_k(\vect{a}_k), \quad \vect{a}_k \in \mathbb{R}^{r_k}, \quad \vect{a}_k(0) = \matr{V}_k^\top(\vect{u}_0-\vect{c}_k).
\end{equation}
The advantage of the ql-ROM approach is that the dynamics are modelled locally, which means that as the trajectory moves along the state space, the cluster affiliations change. The coordinate transformation for switch from cluster $i$ to cluster $j$ is given by
\begin{equation}\label{eq:switch}
    \vect{a}_j(t) = \matr{V}_j^\top \matr{V}_i \vect{a}_i(t) + \matr{V}_j^\top (\vect{c}_i-\vect{c}_j).
\end{equation}
For further details about the algorithm, we refer the reader to \cite{colanera_quantized_2025,Colanera2025b}.
\subsection{Adjoint of ql-ROM}\label{sec:adjoint}
Given a model of the system's dynamics, adjoint-based optimization relies on a direct--adjoint loop: the system is integrated forward in time, the dynamics are linearized along the resulting trajectory, and the corresponding adjoint equations are integrated backward to evaluate gradients~e.g.,~\cite{gunzburger_perspectives_2002,magri2019AdjointMethodsDesign}.

Consider a scalar objective function
\begin{equation}
    \mathcal{J}=\tilde{\mathcal{J}}(\vect u(T)),
\end{equation}
and the ql-ROM dynamics \eqref{eq:dyn_a}. Our goal is to compute the sensitivity of the objective function to the initial condition, i.e., to compute the gradient $d\mathcal{J}/d\vect{u}_0$. Since $\vect a_0=\matr V_{k(0)}^\top(\vect u_0-\vect c_{k(0)})$, it is sufficient to compute $d\mathcal{J}/d\vect a_0$ and then map it to $d\mathcal{J}/d\vect u_0$ by a constant transformation.
We define the Lagrangian
\begin{equation}\label{eq:lagrangian}
        \mathcal{L} = \mathcal{J} - \left\langle \vect{a}^+, \dot{\vect{a}} - \vect{g}(a) \right\rangle = \tilde{\mathcal{J}}(\vect{a}(T)) - \int_0^T \vect{a}^{+\top}\left(\dot{\vect{a}} - \vect{g}(\vect{a})\right) \, dt,
\end{equation}
where $\vect a^+$ are the adjoint variables. Stationarity of $\mathcal{L}$ with respect to $\vect{a}_0$ yields, within each active cluster $k$,
\begin{equation}\label{eq:adjoint}
    \dot{\vect a}_k^+=-\left(\frac{d\vect g_k(\vect a_k)}{d\vect a_k}\right)^\top \vect a_k^+,
\end{equation}
with terminal condition $\vect a^+(T)=d\tilde{\mathcal{J}}/d\vect a(T)$. The term $d\vect{g}_{k}(\vect{a}_{k})/d\vect{a}_{k} \in \mathbb{R}^{r_{k} \times r_{k}}$ in Eq. \eqref{eq:adjoint} is the Jacobian of the local reduced-order model.

When the direct trajectory switches from cluster $i$ to cluster $j$ at time $T_s$, the adjoint satisfies the corresponding jump/coordinate transformation
\begin{equation}\label{eq:switch_cond}
    \vect a_i^+(T_s^-)=\matr V_i^\top \matr V_j\,\vect a_j^+(T_s^+),
\end{equation}
consistent with the direct coordinate change \eqref{eq:switch}. This transformation follows from splitting the integral in Eq. \eqref{eq:lagrangian} at the switching time, and collecting the boundary terms resulting from integration by parts of each subinterval. The reduced gradient follows as
\begin{equation}\label{eq:sensitivityqlrom}
    \frac{d\mathcal{J}}{d\vect a_0}=\vect a^+(0),
\end{equation}
and $d\mathcal{J}/d\vect u_0$ is obtained via the fixed mapping $d\vect a_0/d\vect u_0$.

In this derivation we assume that the switching time is insensitive to the initial condition, i.e.\ $dT_s/d\vect a_0\approx 0$. We verify this assumption in Figure  \ref{fig:init_sensitivity_error}(a). Accounting for event-time sensitivity would require a differentiable switching mechanism and additional terms~\cite{DiBernardo2008,Kong2024Saltation}.

\subsection{Variational data assimilation with ql-ROMs}\label{sec:var_da}
We consider an application in variational data assimilation ~e.g.,~\cite{zaki_turbulence_2025}, where we plug the adjoint of the ql-ROM developed in the previous section in a gradient-based optimization scheme. We consider the problem of inferring the unknown initial condition $\vect{u}_0$ of a trajectory given some observations at the final time $T$. We want to minimize the objective function
\begin{equation}
    \mathcal{J} = ||\vect{y}(T) - \matr{H}\vect{u}(T)||_2^2,
\end{equation}
where $\vect{y} \in \mathbb{R}^{p}$ are the observations and $\matr{H} \in \mathbb{R}^{p \times N}$ is the observation matrix that maps the full state to the observations space. (In this formulation, uncertainties  can be captured by using a covariance matrix and bias in the objective function e.g.,~ \cite{Novoa2024a}.) Starting with an initial guess for $\vect{u}_0$, we propagate the ql-ROM's equations \eqref{eq:dyn_a} forward in time. We linearize the ql-ROM around this trajectory, i.e., compute the local Jacobian at each time step, and integrate the adjoint equations \eqref{eq:adjoint} backward in time, obtaining $d\mathcal{J}/d\vect{u}_0$. We update the initial condition using steepest gradient descent
\begin{equation}
    \vect{u}_0^{(i+1)} = \vect{u}_0^{(i)} - \alpha \frac{d\mathcal{J}}{d\vect{u}_0}^{(i)},
\end{equation}
where $i$ is the iteration step and $\alpha$ is the step size. The procedure is repeated until convergence. We perform the optimization in the full space by updating $\vect{u}_0$ (rather than the reduced coordinates $\vect{a}(0)$). This allows the cluster-affiliation map to be re-evaluated at each iteration, so that the sequence of active local models can adapt during the optimization and the algorithm can converge to the correct trajectory.

\section{Application to spatio-temporal chaos}\label{sec:application_ks}
We demonstrate the developed methodology on a prototypical spatiotemporally chaotic system. The Kuramoto-Sivashinsky (KS) equation is a 1D PDE that arises in a variety of physical phenomena including flame front instabilities~\cite{sivashinsky_flame_1980} and chemical reaction-diffusion equations~\cite{kuramoto_diffusion-induced_1978}. The equation is described by
\begin{equation}
    \frac{\partial u}{\partial t} + u\frac{\partial u}{\partial x} + \frac{\partial^2 u}{\partial x^2} + \nu\frac{\partial^4 u}{\partial x^4} = 0
    \label{eq:ks_pde}
\end{equation}
where $u(x, t): [0, L) \times [0, \infty) \rightarrow \mathbb{R}$ is the velocity on a periodic domain of length $L$, i.e., $u(x, t) = u(x+L, t)$. Letting $\nu = 1$, the dynamical behaviour of the system is determined by the domain length $L$. Following a spectral discretization with 128 Fourier modes, the time integration is performed by a fourth-order exponential time differencing Runge-Kutta method (ETDRK4) \cite{cox_exponential_2002} with a time step of $\Delta t = 0.05$. For the adjoint equations, we employ a discrete time formulation of the equations derived in Section \ref{sec:adjoint}. 

For data generation, we choose $L = 20\pi$, which results in a chaotic regime with a Lyapunov  (LT) of approximately 11.8 time units (235 time steps). The system is initialized by setting $u(x, 0) = \cos(x)$. We discard the first $5 \times 10^3$ time steps as transient, and we utilize the next $10^5$ time steps in training and the next $10^4$ time steps in testing. Following the analysis of \cite{colanera_quantized_2025}, we set the number of clusters $K = 10$ and build local ROMs using $r_k = 30$ POD modes for each cluster. 

\begin{figure}[t]
    \centering
    \includegraphics[width=\linewidth]{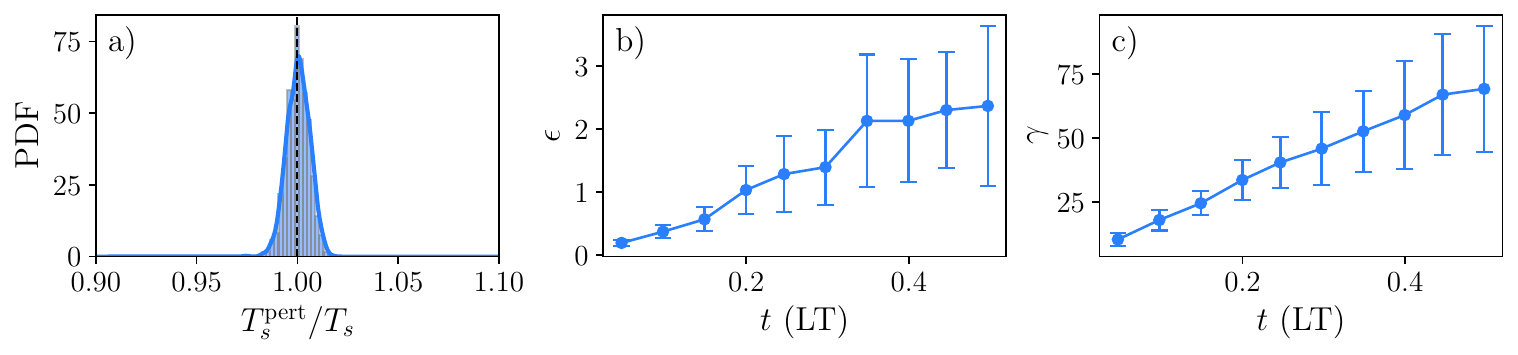}
    \caption{Sensitivity of the switching time and accuracy of the adjoint sensitivity computed using the quantized local reduced model in comparison to the full-order model. (a) Estimated PDF of the normalized $T_s$ in the ql-ROM, by perturbing the initial condition with random noise with  standard deviation $10^{-3}$. (b) the relative $\ell_2$-norm of the error between these two gradient vectors, denoted by $\epsilon$, and (c) the cosine similarity between these two gradient vectors, denoted by $\gamma$. The error bars indicate the mean and $\pm 1$ standard deviation of 50 randomly initialized runs within the test set.}
    \label{fig:init_sensitivity_error}
\end{figure}
First, we verify the accuracy of the sensitivity computation using the ql-ROM. For this purpose, we use an objective function $\mathcal{J} = ||\vect{u}(T)||_2^2$, which is the $\ell_2$-norm of the final state. Figure \ref{fig:init_sensitivity_error} reports (a) the probability density function (PDF) of the normalized $T_s$ obtained from randomly perturbed initial conditions, the accuracy of the ql-ROM adjoint sensitivity $d\mathcal{J}/d\vect{u}_0$ with respect to the full-order model one via two error metrics (b) the relative $\ell_2$-norm of the error between these two gradient vectors, denoted by $\epsilon$, and (c) their cosine similarity, denoted by $\gamma$. Since gradient-based optimization primarily depends on the search direction, the cosine similarity is particularly informative in practice. 
\begin{figure}[t]
    \centering
    \includegraphics[width=\linewidth]{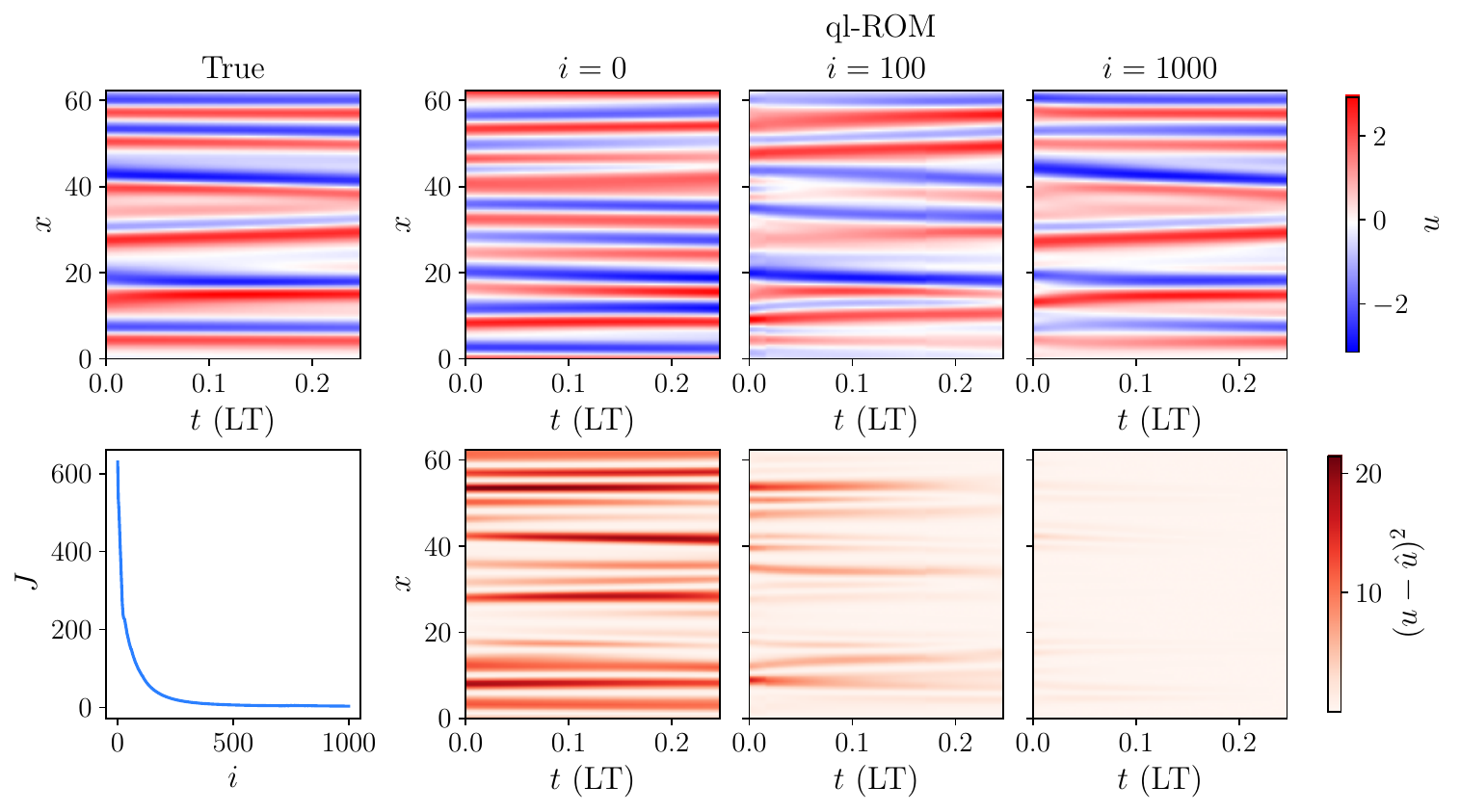}
    \caption{Variational data assimilation on the Kuramoto-Sivashinsky equation using the adjoint of the quantized local reduced-order model. The method reconstructs the true trajectory from measurements at the final time $T = 0.25$ LT following the convergence of the optimization objective, $\mathcal{J}$.}
    \label{fig:var_da}
\end{figure}
As expected in chaotic systems, for short time horizons $\sim 0.25$ LT, the gradient remains accurate, while it gradually loses accuracy as the time horizon increases.

Second, we consider the variational data assimilation problem described in \ref{sec:var_da} for trajectories of length $T = 0.25$ LT. We assume full state measurements, i.e., the observation operator is the identity matrix $\matr{H} = I_{N \times N}$. The ql-ROM is initialized from a random snapshot on the chaotic attractor, and the initial condition is updated via gradient descent using the adjoint-based gradient. The results are shown in Figure \ref{fig:var_da}, where we plot the predicted trajectory at the beginning of the optimization, after 100 iterations, and at the end of the optimization after 1000 iterations. The method successfully recovers the true trajectory with decreasing error near the measurement time. Furthermore, the ql-ROM methodology is significantly more computationally efficient compared to the full-order model. Adjoint-based optimization with the full-order model takes approximately 26 seconds for 100 iterations, whereas with the model reduction enabled by the ql-ROM, this computation time is reduced to approximately 7.4 seconds for 100 iterations, overall resulting in $\times 3.5$ speed-up (results obtained on Intel Xeon(R) Gold 5218R CPU at 2.10 GHz × 80). We expect the cost of the adjoint method to scale with the cost of running two forward simulations. Reducing the state dimension from 128 to 30 would ideally yield a speed-up of about $\times 4.3$, which, in our numerical experiments, is reduced to $\times 3.5$ due to the overhead associated with the cluster switching operations.

\section{Conclusion}\label{sec:conclusion}
We introduced a computationally efficient reduced-order framework for adjoint-based optimization of spatiotemporally chaotic systems. 
 The proposed method combines quantized local reduced-order modelling with adjoint-based optimization at a fraction of the full-order cost.
 We demonstrated the methodology on a variational data assimilation problem for the chaotic Kuramoto-Sivashinsky equation, showing that it can successfully reconstruct the full trajectory over horizons up to 0.25 Lyapunov times from full state measurements at the final time. The proposed algorithm achieves an overall speed-up of $\times 3.5$ compared to the full-order model. In this paper, we used K-means for clustering and obtained the dynamics of the local reduced-order variables using Galerkin projection of the original PDE on local proper orthogonal decomposition modes. These choices are convenient but not essential: alternative clustering strategies and reduced modelling paradigms can be incorporated within the same quantized local and adjoint-based workflow. This method opens up new possibilities for efficient optimization, control, and data assimilation of spatiotemporally chaotic systems.
\begin{credits}
\subsubsection{\ackname} The authors acknowledge funding from the ERC Starting Grant No. PhyCo 949388.
\vspace{-3pt}
\subsubsection{\discintname}
The authors have no competing interests to declare that are
relevant to the content of this article.
\end{credits}
%
%
%

\bibliographystyle{splncs04}
\bibliography{bibliography}
\end{document}